\documentclass[twocolumn,
showpacs,amssymb,aps,prl,superscriptaddress]{revtex4}

\begin{document}

\title{Quantum Anomaly in Molecular Physics}

\author{Horacio E. Camblong}
\affiliation{Department of Physics, University of San Francisco, San
Francisco, California 94117-1080}

\author{Luis N. Epele}
\affiliation{Laboratorio de F\'{\i}sica Te\'{o}rica,
 Departamento de F\'{\i}sica,
Universidad Nacional de La Plata,
\\
 C.C. 67 -- 1900 La Plata, Argentina}

\author{Huner Fanchiotti}
\affiliation{Laboratorio de F\'{\i}sica Te\'{o}rica,
 Departamento de F\'{\i}sica,
Universidad Nacional de La Plata,
\\
 C.C. 67 -- 1900 La Plata, Argentina}

 \author{Carlos A. Garc\'{\i}a Canal}
\affiliation{Laboratorio de F\'{\i}sica Te\'{o}rica,
 Departamento de F\'{\i}sica,
Universidad Nacional de La Plata,
\\
 C.C. 67 -- 1900 La Plata, Argentina}

\begin{abstract}
The interaction of an electron with a polar molecule is shown to
be the simplest realization of a quantum anomaly in a physical
system. The existence of a critical dipole moment
for electron capture and formation of anions, which has been
confirmed experimentally and numerically, is derived. This
phenomenon is a manifestation of the anomaly associated with quantum symmetry
breaking of the classical scale invariance exhibited by the
point-dipole interaction. Finally, analysis of symmetry breaking for this
system is implemented within two different models: point dipole
subject to an anomaly and finite dipole subject to explicit
symmetry breaking. 
\end{abstract}


\maketitle


In this Letter we establish the existence of a remarkably simple
physical realization of a quantum anomaly in Nature. An
anomaly~\cite{tre:85,don:92} is one of the three possible types of
symmetry breaking exhibited by a physical system with an
invariance of some kind---the other two being explicit and
spontaneous symmetry breaking~\cite{hol:00}. In particular, it
arises when a classical invariance of a system is violated upon
quantization. Its physical realizations in Nature have been
recognized in high-energy physics since the introduction of the
Adler-Bell-Jackiw anomaly~\cite{a-b-j1,a-b-j2,a-b-j3}, which
amounts to violation of chiral symmetry. As a practical tool, this
concept has become useful for the analysis of elementary
particles in the standard model~\cite{don:92} and its
extensions~\cite{extensions}, as well as in string
theory~\cite{string}. In addition to its phenomenological
relevance, the study of anomalies has been the source of numerous
theoretical investigations~\cite{tre:85}; for example: (i) in the
path-integral formulation, the chiral-symmetry-breaking anomaly
is due to lack of invariance of the functional
measure~\cite{fuj:79}; (ii) in the Hamiltonian formulation, an
operator becomes anomalous when it does not keep invariant the
domain of definition of the Hamiltonian~\cite{est:86}.

The novel physical realization of a quantum anomaly
that we now consider occurs in the realm of molecular physics.
Specifically,
while the consequences implied by our analysis have been
known in the literature for some time now, their 
interpretation in terms of the anomaly concept is totally new.
Therefore, in this Letter,
by studying this particular realization,
we highlight the relevance and simplicity
of the anomaly phenomenon beyond its original high-energy physics context.
In fact, this remarkable interpretation is the central result of our Letter.

Our analysis involves three essential
ingredients: (i) the identification of the relevant symmetry that
undergoes an anomalous breaking; (ii) the renormalization, 
including dimensional
transmutation~\cite{col:73,cam:00b}, of a nonrelativistic inverse
square potential; and (iii) the subsequent application of the
inverse square potential to the interaction of an electric dipole
with a charge.

 First, the symmetries relevant to our work
are identical to those of the inverse square potential; under time
reparametrizations, they include scale invariance~\cite{jac:72},
which is part of a larger conformal symmetry with SO(2,1) group-theory 
structure~\cite{alf:76}. Surprisingly, the relevance of
these symmetry analyses for the dipole potential has remained
unnoticed despite its central physical importance in molecular
physics. Correspondingly, in this Letter, we review and adapt the
symmetry arguments to the case of the dipole-charge interaction.

Second, we have recently considered the inverse square potential
using field-theory techniques and shown the existence of a
critical coupling for the onset of dimensional
transmutation~\cite{cam:00}. On the other hand, dimensional
transmutation has been recently viewed as an example of a quantum
anomaly: that associated with the breaking of scale
invariance~\cite{jac:91,hol:93}. However, this anomaly has been
discussed only in the context of the two-dimensional
delta-function potential~\cite{cab:98}.
 In this
Letter we show that the concepts of anomaly and dimensional
transmutation apply to the inverse square potential and related
physical problems, including the intriguing electric dipole-charge
interaction, with far-reaching implications in molecular physics.

Third, the nonrelativistic interaction of an electric dipole
with a charge poses a problem whose physical relevance was first
recognized in nuclear physics~\cite{fer:47,wig:49} and then in
molecular physics~\cite{lev:67,bro:67,cra:67}. In the traditional
treatments of this problem~\cite{tur:77}, a key ingredient is the
existence of a critical dipole moment for the capture of the
electric charge by the dipole. This unambiguous prediction of
quantum mechanics has led to a standard lore in molecular
physics, according to which any neutral molecule with a dipole
moment of the order of 2 D or greater~\cite{debye}
 should be capable of
capturing an electron and sustaining a molecular
anion~\cite{cra:71}. Recent experiments~\cite{mea:84,des:94} and
numerical simulations~\cite{des:94} confirm this generic
prediction. This immediately leads one to pose the question: what
is the origin of this critical coupling and why is it robust? In
this Letter we shed light on this issue by treating the problem
using scale invariance and its associated anomaly, dimensional
transmutation.

The interaction of an electric charge $Q$ with a point dipole
${\bf p}$, which we treat nonrelativistically and refer to as the
point-dipole potential, is described by means of
\begin{equation}
V({\bf r})= K_{e} \, \frac{ Q \, p \, \cos \theta}{r^{2}} 
\;  ,
\label{eq:point_dipole_potential}
\end{equation}
which can be regarded as a generalized inverse square potential
with an anisotropic coupling strength. As usual, in
Eq.~(\ref{eq:point_dipole_potential}), the polar angle $\theta$ is
measured from the direction of the dipole moment, and we assume
that the problem occurs in ordinary three-dimensional space. The
coupling can be rewritten in a  dimensionless form
\begin{equation}
\lambda = - \frac{2 m \, K_{e}}{\hbar^{2}}{ \,  Q \, p } =
\frac{p}{p_{0}} 
\; , 
\label{eq:coupling}
\end{equation}
with $m$ being the reduced mass of the system and $K_{e}$ the
electrostatic constant. In Eq.~(\ref{eq:coupling}) the
characteristic dipole moment $p_{0}$ sets the scale for our
analysis and is the relevant dimensional parameter whose order of
magnitude will provide an estimate for criticality (charge
capture). For the particular case of an electron interacting with
a polar molecule, $Q=-e$ (with $e$ being the electron charge magnitude), 
and $p_{0}=e a_{0}/2 \approx 1.271 D$, 
where $a_{0}$ is the Bohr radius and $D $ is the
debye~\cite{debye}. The Hamiltonian associated with
Eqs.~(\ref{eq:point_dipole_potential}) and
(\ref{eq:coupling}),
\begin{equation}
H = - \frac{\hbar^{2}}{2m} \left[
 \nabla^{2}
+ \lambda \, \frac{\cos \theta}{r^{2}}
\right]
\; ,
\label{eq:Hamiltonian}
\end{equation}
is explicitly scale and conformally invariant,
as the analysis below shows.

The corresponding classical Lagrangian $L = m v^{2}/2
 - V({\bf r})
 $ has an associated action that is
  invariant under the scale
transformations $t \rightarrow \tau t$, ${\bf r} \rightarrow
\varrho {\bf r}$, (with $\tau>0$ and $\varrho>0$), $\varrho^{2}=
\tau$; this property is shared by the larger class of homogeneous
potentials of degree $-2$~\cite{cam:00b}, which also includes the
two-dimensional delta-function potential. In the language of
dimensional analysis, this symmetry means that the point-dipole
potential has no characteristic dimensional scales and the
coupling $\lambda$ is dimensionless~\cite{self-adjointness}.

The symmetry analysis under time reparametrizations can be
generalized to~\cite{jac:80,jac:90} $t \rightarrow \tilde{t} = t -
\alpha f(t) $, ${\bf r} \rightarrow \tilde{\bf r} (\tilde{t}) =
J^{\delta} \, {\bf r}(t)$, with $J= |d \tilde{t}/dt|$. Invariance
of the action occurs only when the following two conditions are
simultaneously satisfied: (i) $\delta=1/2$ and (ii) $f(t)$ is
quadratic in $t$. This selects the SO(2,1) conformal group, just
as for the inverse square potential~\cite{alf:76}, the magnetic
monopole~\cite{jac:80}, and the magnetic vortex~\cite{jac:90},
with the following three generators: (i) the Hamiltonian $H$,
Eq.~(\ref{eq:Hamiltonian}), associated with time translations $t
\rightarrow t - \alpha $
 [for $f(t) =1$];
(ii) the dilation generator
\begin{equation}
D= t H - \frac{1}{4} \left( {\bf r} \cdot {\bf p} + {\bf p} \cdot
{\bf r}
 \right)
 \;  ,
\end{equation}
associated with the scale transformation defined in the previous
paragraph, with $\tau= 1 - \alpha$ [for $f(t)=t$]; and (iii) the
conformal generator
\begin{equation}
K= H t^{2} - \frac{1}{2} ({\bf p} \cdot {\bf r} + {\bf r} \cdot
{\bf p} ) \,
 t + \frac{1}{2} m
r^{2}
 \;  ,
\end{equation}
associated with the time special conformal transformation $1/t
\rightarrow 1/t + \alpha$ [for $f(t)=t^{2}$].
 The corresponding
commutators $[D,H]= -i \hbar H$, $[D,K]= i \hbar K$, and $[H,K]=
2 i \hbar D$ show that these three operators form an SO(2,1)
algebra~\cite{wyb:74}.

We will now examine how scale invariance is broken at the
quantum-mechanical level. This symmetry breaking manifests itself
in the appearance of a critical dipole moment, whose existence
and numerical value we will establish next by generalizing our
treatment of the inverse square potential~\cite{cam:00}. This goal can
be accomplished by writing the Schr\"{o}dinger equation for the point
dipole in spherical coordinates, with the separation of variables
 $\Psi (r,\theta, \phi)= u(r) \Theta (\theta) e^{im\phi}/r$,
where the azimuthal dependence corresponds to conservation of the
axial component $L_{z}$ of angular momentum. Then, the
corresponding equations for $r$ and $\theta$ are explicitly given
by
\begin{equation}
\frac{d^{2}  u(r)}{dr^{2}} + \left( \eta + \frac{\gamma}{r^{2}}
\right) u(r) = 0 \;
\label{eq:radial_eq}
\end{equation} 
and
\begin{equation}
\hat{A} \Theta (\theta) = \gamma \Theta (\theta) \; ,
\label{eq:theta_eq}
\end{equation}
where $\eta=2mE/\hbar^{2}$,
\begin{equation}
\hat{A} = - \Lambda^{2} + \lambda \cos \theta
 \; ,
\label{eq:angular_operator} 
\end{equation}
and
  $\Lambda^{2} =
L^{2} /\hbar^{2}$ is the dimensionless angular momentum (squared).
Equations~(\ref{eq:radial_eq}) and~(\ref{eq:theta_eq}) constitute
a coupled system of eigenvalue equations linked by the separation
constant $\gamma$. Notice that Eq.~(\ref{eq:radial_eq}) can be
interpreted as defining an isotropic inverse square potential for
the zero angular-momentum channel in three dimensions, with
coupling strength $ \lambda_{\rm ISP} \equiv \gamma$. In addition,
$\gamma$ is implicitly related to the actual coupling $\lambda$ of
the point-dipole potential by means of the eigenvalue
equation~(\ref{eq:theta_eq}). This relation can be obtained by
recasting Eq.~(\ref{eq:theta_eq}) into matrix form through the
matrix
\begin{equation}
M(\gamma, \lambda) = - A(\lambda) + \gamma  \, \openone \; 
\end{equation}
(where $\openone$ is the identity matrix)
and subsequently setting up
the corresponding characteristic 
equation~\cite{lev:67,bro:67,cra:67}
\begin{equation}
D(\gamma, \lambda) \equiv \det M (\gamma,\lambda) 
= 0 
\; .
 \label{eq:determinant}
\end{equation}

It should be noticed that the existence of symmetry breaking for
the dipole potential can be
viewed as a consequence of the corresponding symmetry breaking for
the inverse square potential.
As shown in Ref.~\cite{cam:00}, this quantum anomaly 
occurs for the inverse square potential in the supercritical
or strong-coupling regime $ \gamma \geq \gamma^{(*)}=1/4$
(for $l=0$ in three dimensions).
It then follows that there exists
a critical value $\lambda^{(*)}$
of the dimensionless dipole moment 
to be determined from the corresponding critical inverse-square coupling
$\gamma^{(*)}=1/4$, i.e.,
\begin{equation}
D(\gamma^{(*)},\lambda^{(*)}) =0 \; ,
\label{eq:critical_characteristic_equation}
\end{equation}
and such that a quantum anomaly occurs for 
the strong-coupling regime $ \lambda \geq \lambda^{(*)}$.
A straightforward calculation
shows that 
$\lambda^{(*)} \approx 1.279$,
which amounts to the familiar critical dipole moment $p^{(*)} \approx
1.625 D$~\cite{fer:47,wig:49,lev:67,bro:67,cra:67}. Extensive
empirical and numerical studies have confirmed the existence of a
critical dipole moment of a similar value for a large number of
molecules~\cite{mea:84,des:94}. This remarkable universal property
of polar molecules can be regarded as the simplest physical
example of a quantum anomaly.

The fact that the predictions arising from the quantum anomaly
analyzed in the previous paragraph agree with the corresponding
empirical and numerical findings requires further elaboration. In
effect, a polar molecule is better modeled as a finite dipole,
with an interaction potential
\begin{equation}
V({\bf r}) = K_{e} Qq \, \left( \frac{1}{R_{+}} - \frac{1}{R_{-}}
\right) =
 K_{e} \, \frac{ Q \, p \, \cos \theta}{r^{2}}
 +
 V_{\rm sb}({\bf r})
\;  .
\label{eq:exact_finite_dipole_potential}
\end{equation} 
In Eq.~(\ref{eq:exact_finite_dipole_potential})
$R_{\pm}$ represents the distance to the charge $Q$ from the positive
and negative charges of the dipole, which are separated by a distance $a$.
Equation~(\ref{eq:exact_finite_dipole_potential}) displays the
point-dipole potential~(\ref{eq:point_dipole_potential}), with
$p=qa$ ($l=1$), supplemented by a symmetry-breaking potential
$V_{\rm sb}({\bf r})$, which  includes higher-order multipoles
(for $l>1$ and $r>a/2$, with moments $\sim qa^{l}=p a^{l-1}$), as
well as the contribution to the potential for $r<a/2$. In short,
in this model, the SO(2,1) symmetry of
potential~(\ref{eq:point_dipole_potential}) undergoes explicit
symmetry breaking by the introduction of additional terms in the
Hamiltonian. The corresponding Schr\"{o}dinger equation with
potential~(\ref{eq:exact_finite_dipole_potential}) can be derived
by separation of variables in prolate spheroidal
coordinates~\cite{prolate}, thus providing a
solution~\cite{lev:67,bro:67,cra:67,finite_dipole} that
illustrates the effect of adding explicit symmetry-breaking terms.

{\em A priori\/}, it is by no means obvious that the approximate
point-dipole representation captures the correct behavior and the
correct numerical value of the critical dipole moment. However,
as we will show next, this is indeed the case.
 In other words, even though the finite dipole introduces a length scale $a$
 and amounts to an example of explicit symmetry breaking, the
existence of a critical dipole moment as well as its numerical
value are independent of $a$. In fact, this result is confirmed
by the explicit solution of the problem with
potential~(\ref{eq:exact_finite_dipole_potential}).
 In short, the simplified point-dipole model exhibits an
anomaly, whose relevance is highlighted by a robust
prediction---one that survives when the finite size of the
molecule is considered.

Let us now see the dimensional argument
that proves the statement of the previous paragraph. The
characteristic dimensional parameters for the dynamics of the
finite dipole are $\hbar$, $m$, $q$, and $a$, as well as the
finite charge $Q$; moreover the interaction only involves the
product $Qq$.
 Then, according to Buckingham's Pi theorem of
dimensional analysis~\cite{pi_theorem},
\begin{equation}
E_{\rm gs} = - \frac{\hbar^{2}}{2m a^{2}} \, F (\lambda) \;  ,
\end{equation}
where $\lambda $ is the dimensionless combination of the given
parameters that we previously defined in Eq.~(\ref{eq:coupling})
and $F(\lambda)$ is an arbitrary function of $\lambda$. On the
other hand, the critical value of the dimensionless coupling,
$\lambda^{(*)}$, occurs when the ground state energy $E_{\rm gs}$
comes into existence going through a 
zero value~\cite{energy_subtlety}. 
Thus, the critical dipole moment is defined by
the condition
\begin{equation}
F(\lambda^{(*)}) =0 \;  ,
\end{equation}
whose solution is a dimensionless number independent of the size
$a$ of the dipole. In particular, the critical value survives in
the limit $a \rightarrow 0$, which amounts to the ideal point
dipole. This shows that the point-dipole model with anomalous
symmetry breaking predicts the correct physics of the finite
dipole, for which the symmetry is explicitly broken.

In conclusion, we have found theoretical and empirical
evidence---further confirmed by numerical computations---of the
existence of a quantum anomaly in molecular physics.
Specifically, this anomaly is manifested by the formation of
anions through electron capture by polar molecules with
supercritical dipole moments and, to our knowledge, represents
the simplest realization of quantum-mechanical symmetry breaking
in a physical system.

 This research was supported in part by CONICET and ANPCyT,
Argentina (L.N.E., H.F., and C.A.G.C.) and by the University of
San Francisco Faculty Development Fund (H.E.C.). Instructive
discussions with Profs.\ Carlos R. Ord\'{o}\~{n}ez and B. Montgomery
Pettitt are gratefully acknowledged by H.E.C.


\begin{thebibliography}{99}
\bibitem{debye}
The debye $D$ is a characteristic scale for  molecular dipole
moments, defined to be exactly  $1$ $D = 10^{-18}$ esu-cm.
\bibitem{self-adjointness}
This property implies that, quantum-mechanically and for a
sufficiently strong potential, the Hamiltonian fails to be bounded
from below and is not self-adjoint, despite its Hermitian
character. A relevant example of the application of the technique
of self-adjoint extensions can be found in Ref.~\cite{jac:91},
where it is applied to the two-dimensional delta-function
potential. This point is related to Ref.~\cite{est:86}.
\bibitem{energy_subtlety}
The rapid growth in the value of $|E_{\rm gs}|$
in the molecular system, as $\lambda$ is increased beyond
$\lambda^{(*)}$, 
corresponds to the well-known arbitrariness in the value 
of the anomalous scale in the renormalized theory~\cite{cam:00}.
\end{thebibliography}
\end{document}